\documentclass[%
 aip,
 amsmath,amssymb,
 reprint,%
]{revtex4-1}

\usepackage{graphicx}
\usepackage{dcolumn}
\usepackage{bm}

\usepackage[utf8]{inputenc}
\usepackage[T1]{fontenc}
\usepackage{mathptmx}
\usepackage{etoolbox}
\usepackage{siunitx}
\usepackage{amsmath}  

\makeatletter
\def\@email#1#2{%
 \endgroup
 \patchcmd{\titleblock@produce}
  {\frontmatter@RRAPformat}
  {\frontmatter@RRAPformat{\produce@RRAP{*#1\href{mailto:#2}{#2}}}\frontmatter@RRAPformat}
  {}{}
}%
\makeatother
\begin{document}

\preprint{AIP/123-QED}

\title[Droplet-induced vortex ring formation]{On the vortex ring formation and mixing in thin films upon droplet impact}
\author{Hatim Ennayar}
 \altaffiliation[\textbf{Author to whom correspondence should be addressed:} ennayar@sla.tu-darmstadt.de]{}
 \author{Juan Camilo Dueñas Torres}
 \author{Philipp Brockmann}
 \affiliation{ 
Institute for Fluid Mechanics and Aerodynamics, TU Darmstadt, Flughafenstr. 19, 64347, Darmstadt-Griesheim \\ 
}
 \author{Hyoungsoo Kim}%
\affiliation{ 
Department of Mechanical Engineering, Korea Advanced Institute of Science and Technology, Daejeon \\ 
}%
\author{Jeanette Hussong}%
\affiliation{ 
Institute for Fluid Mechanics and Aerodynamics, TU Darmstadt, Flughafenstr. 19, 64347, Darmstadt-Griesheim \\ 
}%


\date{\today}

\begin{abstract}
When a droplet impacts a thin liquid film, ring vortices are formed that play a key role in momentum and species transport. The present study provides new insights into the vortex ring dynamics during droplet impact onto thin liquid films. More precisely, the dynamics of ring vortex formation, propagation, and decomposition in thin liquid films are investigated experimentally to elucidate the role of vortex-wall interaction on vortex ring instabilities and the formation of different vortex regimes. 
 Using particle image velocimetry (PIV) and laser-induced fluorescence (LIF), the influence of Reynolds number $Re$, Weber number $We$, and dimensionless film thickness $\delta$ on the vortex ring dynamics was studied. Specifically, experiments were conducted in ranges of $Re$ to $3900$, $We$ to $61$, and $\delta$ from $0.09$ to $1.35$, covering thin and thick film regimes. The influence of film thickness on vortex-ring formation and its evolution was studied. A focus is placed on vortex-ring instabilities, which are experimentally observed as a rapid transition from a rotationally symmetric single vortex ring to ring-shaped formations with multiple vortices as the film height decreases.
 A regime map for different ring vortex states, associated with different shaped vortices and instabilites, was derived in $Re$-$\delta$ space for droplet impact on thin liquid films. It was observed that instabilities occur at lower $Re$ for thinner films, while no instability was observed in the thick film regime for $Re$ up to $3900$. The azimuthal wave number of vortex rings (finger-shaped perturbations) was correlated with $Re$ and $\delta$ during the transition. A linear increase in wave number was observed with increasing $Re$, while it decreases with increasing film thickness. It was shown that the circulation strength of primary vortex rings decreases more rapidly for thinner liquid films due to wall interactions. At the same time, the formation of secondary vortex rings can be observed as $Re$ decreases. Based on these measurements, an empirical model is proposed to predict the temporal evolution of total vortex-ring circulation, accounting for both the vortex generation and decay phases.

\end{abstract}

\maketitle

\section{\label{sec:1}Introduction}

Understanding the complex dynamics of droplet impact on liquid surfaces is crucial due to its widespread occurrence in natural phenomena and industrial applications such as spray coating \cite{eslamian2014spray}, cooling \cite{kim2007spray,gajevic2023spray}, or fuel injection systems \cite{schmidt2021near,pati2022numerical}. Mixing during droplet impact on liquid films is of particular interest due to its relevance in applications like fuel injection, where efficient mixing is crucial for combustion efficiency and emission reduction \cite{maliha2022optical}. 

The outcome of a droplet impact on a film is governed by Reynolds number $Re=\frac{UD}{\nu}$, Weber number $We=\frac{\rho DU^2}{\sigma}$ and the dimensionless film thickness $\delta=\frac{h}{D}$. Here, $U$ represents the drops impact velocity, $D$ is the droplet diameter, $\nu$ is the kinematic viscosity, $\rho$ is the density, $\sigma$ is the surface tension, and $h$ is the film thickness. The effect of these parameters on the droplet impact dynamics are extensively discussed in literature \cite{yarin2006drop,josserand2016drop,breitenbach2018drop}. 

Liquid films are commonly classified into four distinct categories denoted as very thin films, thin films, thick films and deep pools \cite{tropea1999impact,liang2016review}. First, thin liquid films are defined by Cossali \textit{et al.}\cite{cossali1997impact} as those with $\delta <1$, whereas \citet{tropea1999impact} extended this classification to include film thicknesses up to $\delta < 1.5$. More recently, \citet{geppert2019experimental} refined the classification, defining thin films as those within the range of $0.08 \leq \delta \leq 0.6$, which will be adopted in this study. In the thin film regime, both the film thickness and wall features significantly influence the impact behavior of a droplet. On the other hand, for thick films, characterized here by $\delta > 0.6$, the impact dynamics are governed primarily by film thickness and are independent of wall features \cite{tropea1999impact}. In contrast, films with $\delta < 0.08$ are considered to be very thin films, where wall effects dominate the outcome of the impact \cite{geppert2019experimental}. Finally, the deep pool regime defined by $\delta \gg 4$ describes conditions where the impact behavior is independent of the film thickness\cite{tropea1999impact}. In this work, for the first time both thin and thick film regimes are analyzed to capture the influence of film thickness on the dynamics on the droplet impact and associated instabilities. 

Several studies have investigated mixing mechanisms during droplet impact. \citet{ersoy2019capillary} identified six different mixing mechanisms, including the expansion and receding of the crater, collapse of the crown on the film, crown-finger formation, secondary droplets, surface waves resulting from the impact, and finally diffusion. Khan \textit{et al.} \cite{khan2020droplet} explored droplet impact on vibrating films, showing that induced vibrations can enhance the mixing efficiency by 50\%. Additionally, Parmentier \textit{et al.}\cite{parmentier2023drop} observed how the initial film thickness and Weber number strongly influence the resulting mixing patterns. Recently, we utilized Laser-Induced Fluorescence (LIF) to quantify species transport during droplet impact onto thin liquid films \cite{ennayar2023lif}. Spatially and temporally resolved concentration fields revealed that vortex rings formed during the droplet impact also in very thin films being a key mechanism that drives mixing. To better understand and predict mixing processes in thin films, it is crucial to elucidate onto the detailed vortex dynamics in thin films upon droplet impact, which is the subject of the present study. 

The formation of a vortex ring during droplet impact arises from the azimuthal vorticity generated by high-velocity gradients at the liquid-liquid interface immediately after impact \cite{cresswell1995drop}. Vortex ring formation has been extensively studied in the context of droplet impacts into deep pools \cite{saha2019kinematics,agbaglah2015drop,thoraval2016vortex,lee2015origin,cresswell1995drop,peck1994three}. However, their dynamics in thin liquid films have not been explored thoroughly, leaving gaps in understanding their behavior and influence in this regime. In thin films, the interaction between the droplet-induced vortex ring and the wall introduces additional complexities to the vortex dynamics. Wilkens et al. \cite{wilkens2013vortex} investigated vortex ring generation during droplet impact on thin films with $\delta = 0.37$ using a top view to study long-wavelength perturbations. Additionally, short-wavelength fingering were observed in more recent work \cite{ennayar2023lif}, leading to chaotic mixing during the impact.

The dynamics of a vortex ring hitting a wall in a semi-infinite domain has been studies by \citet{walker1987impact}. They showed that the vortex ring undergoes radial stretching upon impact with the wall. This stretching leads to boundary layer separation and ultimately results in the formation of a secondary vortex ring with opposite-signed vorticity. This secondary ring orbits the primary vortex ring, potentially leading to the rebound of the primary vortex ring from the wall, as observed in the work of Chu \textit{et al.}\cite{chu1995vortex}, as well as of \citet{harris2012instability}. 
At high energy impacts, the secondary vortex ring exhibits azimuthal instabilities\cite{walker1987impact,cerra1983experimental,cheng2010numerical} that become visible through a circumferential vortex waviness. Two primary types of instabilities are responsible for vortex ring instability \cite{archer2010instability}. Long-wavelength perturbations, known as Crow instability \cite{crow1970stability}, and short-wavelength perturbations, referred to as Tsai-Widnall-Moore-Saffman (TWMS) or elliptical instability \cite{tsai1976stability,moore1975instability}. The ratio between vortex circulation $\Gamma$ and the kinematic viscosity $\nu$ is usually referred to as circulation Reynolds number $Re_\Gamma = \Gamma / \nu$. At lower  $Re_\Gamma$, the Crow instability governs the vortex-wall interaction dynamics. At higher $Re_\Gamma$, the elliptical instability dominates, leading to the breakdown of the vortex rings into a turbulent cloud \cite{archer2010instability}. These instabilities have been observed in both experimental \cite{walker1987impact,leweke1998cooperative,harris2012instability,mckeown2020turbulence} and numerical studies \cite{swearingen1995dynamics,archer2010instability,cheng2010numerical,mishra2021instability} for vortex ring collisions with walls. In thin liquid films, however, the dynamics differ due to the limited film thickness. The interaction between the vortex ring and the wall is influenced by the finite fluid volume, potentially altering the mechanisms leading to instabilities. To date, investigations of vortex ring dynamics in thin liquid film during droplet impact have not been studied in detail.

To close this gap, for the first time a comprehensive study is conducted on the vortex ring dynamics during droplet impact on liquid films ranging from thick films down to thin films. Particle Image Velocimetry (PIV) and Laser-Indiced Fluorescence (LIF) are utilized to study the influence of $Re$, $We$, and $\delta$ on the vortex ring dynamics.

\section{\label{sec:2}Experimental setup}
\begin{figure}[htbp]
    \centering
    \includegraphics{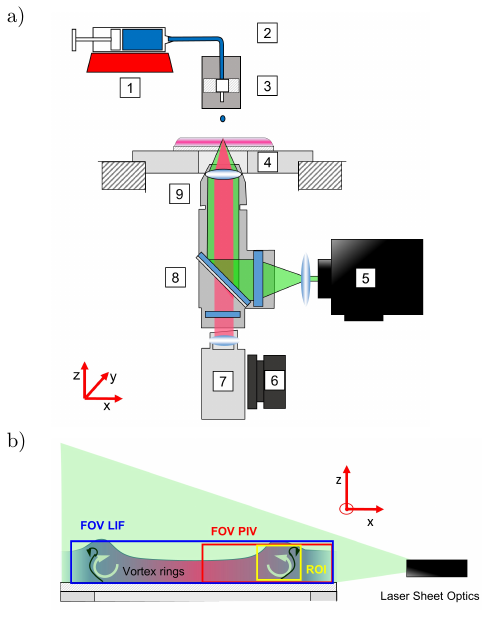}
       \centering
    \caption{\textbf{a)} Schematic illustration of bottom view setup: (1) Syringe pump, (2) z-Traverse, (3) Cannula, (4) Thin liquid film on FTO glass substrate, (5) High power LED, (6) x,y,z-Traverse, (7) HS-Camera, (8) Dichroic mirror with bandpass filters, (9) Microscope Objective. \textbf{b)} Schematic illustration of side view setup (HS - camera orientated along y- axis facing x-z plane. The blue and red boxes represent the field of view during LIF and PIV measurements, respectively. The yellow box represent the region of interest during PIV measurements}
    \label{fig:1}
\end{figure}

The experimental investigation of vortex ring dynamics during droplet impact on thin liquid films was conducted using two primary configurations: a bottom view setup and a side view setup, as illustrated in Fig. \ref{fig:1}a and Fig. \ref{fig:1}b, respectively. Each setup is designed to capture specific aspects of the impact process and comprises three main components: the drop generator, the impact substrate, and the optical system.

The drop generator system consists of a $\SI{5}{\milli\litre}$ syringe (Braun GmbH) driven by a precision syringe pump (Aladdin AL-1010, WPI). The syringe is connected via a tube to a blunt needle (Braun GmbH), which is mounted on an automatically controlled traverse allowing precise adjustments of the drop height. A constant flow rate is maintained to generate repeatable water droplets with diameters of $D = 2.225 \pm \SI{0.025}{\milli\meter}$ and $D = 3.9 \pm \SI{0.025}{\milli\meter}$.

\begin{figure*}
\centering
\includegraphics{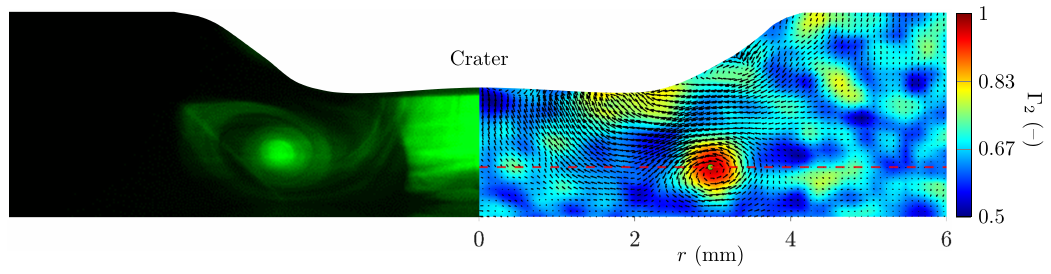}
  \caption{(Left) Visualization of the vortex ring during droplet impact on liquid film using Laser-Induced Fluorescence. (Right) Velocity vector field overlaid with $\Gamma_2$ scalar field extracted from PIV measurements. The green point indicates the center of the vortex. The case corresponds to $\delta = 0.90$, $We = 54$, $Re = 3900$ at $t=\SI{27}{\milli\second}$.}
    \label{fig:2}
\end{figure*}

\subsection{Bottom view}
For recordings from a bottom view, a \SI{50}{\milli\meter} $\times$ \SI{50}{\milli\meter} glass plate (Sigma-Aldrich) is used as substrate. To enhance surface energy for better adhesion of the liquid film, the substrate is thoroughly cleaned with water and detergent, followed by immersion in an ultrasonic bath of isopropanol. The substrate is then rinsed thoroughly with water to remove any remaining isopropanol. Deionized water mixed with a fluorescent dye is spread across the substrate. The film thickness is adjusted by adding or removing small amounts of liquid using a syringe until the desired thickness is achieved. The initial film thickness is precisely measured using a chromatic-confocal point sensor (confocalDT IFS2407-0.8, Micro-Epsilon) with an accuracy of $\pm \SI{0.4}{\micro\meter}$. The sensor is mounted on a motorized arm, enabling a three-dimensional displacement to the sensor to scan the overall film thickness. The thicknesses $h$ are chosen to achieve dimensionless thicknesses $\delta = \frac{h}{D}$ ranging from 0.09 to 1.35, as shown in Table \ref{tab:1}, along with other initial parameters. Furthermore, the initial parameters were carefully selected to ensure that the droplet impact outcomes remain within the deposition regime, thereby avoiding complexities introduced by corona collapse or splashing phenomena. This approach isolates the vortex ring dynamics and their influence on mixing.

The bottom view observations are conducted using a custom-built microscope with an optical tube (InfiniTube Special, Infinity Photo-Optical), illuminated by a $\SI{7}{\watt}$ high-power green LED ($\lambda \approx \SI{532}{\nano\meter}$, ILA iLA.LPS v3) \cite{ennayar2023lif,brockmann2022utilizing,brockmann2025enhancement}. The optical path includes a dichroic mirror and band-pass filters (Thorlabs DFM1/M) to direct the excitation light through the objective and filter the fluorescence emission. An Infinity Photo-Optical IF-3 objective lens with a magnification of $1\times$ is used. High-speed imaging is performed with a 12-bit, 2048 $\times$ 1952 pixels CMOS camera (Phantom T1340, Vision Research). The entire apparatus can be traversed in three dimensions with step sizes of $\SI{1.25}{\micro\meter}$, allowing for precise positioning.

Rhodamine B dye (Carl Roth) is selected as the fluorescent tracer in the bottom view experiments due to its advantageous spectral properties, with an absorption maximum around $\SI{550}{\nano\meter}$ and an emission maximum at $\SI{590}{\nano\meter}$. The large Stokes shift facilitates the separation of excitation and emission wavelengths, which is critical for the single optical access used for both illumination and imaging. One noted disadvantage of Rhodamine B is its influence on the surface tension of water depending on its concentration \cite{seno2001}. However, at the concentration used in this study ($C = 8 \times 10^{-5} \mathrm{M}$), no significant effect on surface tension is observed, as demonstrated in our previous work \cite{ennayar2023lif}.

\begin{figure}[htbp]
    \centering
    \includegraphics{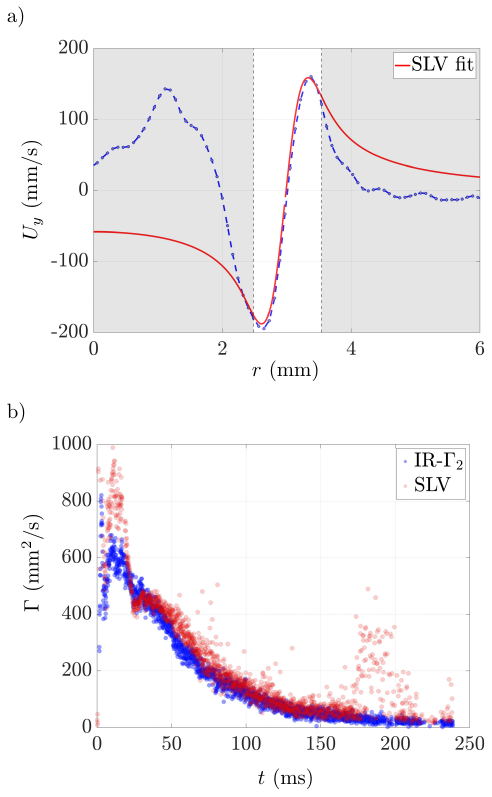}
       \centering
    \caption{\textbf{a)} Least-squares fit of azimuthal velocities of two superposed Lamb-Oseen vortices (red line) and tangential velocity profile measured along a horizontal line passing through the centers of the vortex pair for $\delta=0.90$, $We=54$ and $Re=3900$ at $t=\SI{27}{\milli\second}$ (blue dots). White shaded area represents the boundary of the vortex used for the fit. \textbf{b)} Time evolution of vortex circulation for the case $\delta=0.90$, $We=54$ and $Re=3900$ calculated by Stokes' Theorem (blue) and SLV method (red).}
    \label{fig:3}
\end{figure}

\begin{table}
\caption{\label{tab:1}Water properties and impact parameters.}
\begin{ruledtabular}
\begin{tabular}{lc}
Parameters and fluid properties & Value\\
\hline
Kinematic viscosity  $\nu$ (\SI{}{\milli\square\meter/\second}) & 1.004\\
Density  $\rho$ (\SI{}{\kilogram/\cubic\meter}) & 997\\
Surface tension  $\sigma$ (\SI{}{\kilogram/\square\second}) & 0.0723\\
Film thickness $h$ (\SI{}{\micro\meter}) & 100,...,5265\\
Dimensionless thickness $\delta$ (--) & 0.09,...,1.35\\
Droplet diameter (\SI{}{\milli\meter}) & 2.225; 3.900\\
Reynolds number $Re$ (--) & 1400,...,3900\\
Weber number $We$ (--) & 7,...,61\\
\end{tabular}
\end{ruledtabular}
\end{table}

\subsection{Side view}
The same droplet generator is used for both the side and bottom views. A custom-made cuvette of dimensions \SI{50}{\milli\meter} $\times$ \SI{50}{\milli\meter} $\times$ \SI{50}{\milli\meter} made of acrylic is used as substrate. The surface of the cuvette is treated with a BD-10AS high-frequency generator (Electro-Technic Products) to enhance wettability. Side view recordings are captured with a high-speed camera (Photron Fastcam Mini-AX200). Illumination is provided by a $\SI{532}{\nano\meter}$ green laser (Microvec) equipped with sheet optics to create a thin laser sheet passing through the center of the impact region.

Two different measurement techniques are combined here: Laser-Induced Fluorescence (LIF) for visualization of drop-film fluid engulfment and Particle Image Velocimetry (PIV) to identify vortices and to quantify the evolution of their circulation strength. For LIF measurements, the droplets are dyed with Rhodamine 6G (Carl Roth), which does not significantly affect the surface tension of water at the concentrations used \cite{seno2001}. For PIV measurements, $\SI{10}{\micro\meter}$ glass hollow spheres (LaVision GmbH) are added to both the droplet and the liquid film to serve as tracer particles. The field of view (FOV) for both LIF and PIV measurements is illustrated in Fig. \ref{fig:1}b. 

All experiments were repeated at least five times for each set of conditions to ensure reproducibility of the results.


\section{\label{sec:3}Method}
To investigate the spatial and temporal evolution of vortex rings during droplet impact on thin liquid films, PIV measurements were performed. Depending on the impact velocity and the film thickness, the high-speed camera operated at three different frame rates: 10,000 fps, 8,100 fps, and 6,400 fps. The captured images were processed using \textit{LaVision DaVis 10.2} software, with interrogation windows of 24 $\times$ 24 pixels and 75\% overlap. Figure \ref{fig:2} illustrates both the LIF recording and the PIV results of the vortex ring during impact at $t=\SI{27}{\milli\second}$ for the case $\delta=0.90$, $We=54$, $Re=3900$. 
Due to the light sheet sectioning, the vortex ring is visualized as a vortex pair. For the PIV analysis, only one side of the vortex pair was specifically focused on to achieve better spatial resolution (see red FOV in Fig. \ref{fig:1}). This detailed focus is essential since accurately identifying the vortex region is crucial for extracting vortex circulation. This latter is a key for understanding vortex behavior during droplet impact on thin liquid films. To this end, a Galilean invariant scalar field $\Gamma_2$ was employed \cite{graftieaux2001combining}. This approach provides a reliable means to identify vortex boundaries and core locations \cite{kissing2021delaying}. As described by Graftieaux \textit{et al.}\cite{graftieaux2001combining}, the $\Gamma_{2}$ function is defined as:

\begin{eqnarray}
\Gamma_2(P) = \frac{1}{S} \int_S \frac{[\vec{PM} \times (\vec{U}_M - \vec{\tilde{U}}_P)] \cdot \vec{z}}{||\vec{PM}|| \cdot ||\vec{U}_M - \vec{\tilde{U}}_P||} \, dS
    \label{Eq:1}
\end{eqnarray}

where $S$ is a two-dimensional area surrounding point $P$, $M$ is a point within $S$, and $\vec{z}$ is the unit vector normal to the measurement plane. The vector $\vec{PM}$ represents the position vector from $P$ to $M$, 
$\Vec{U}_{M}$ is the velocity vector at point 
$M$, and $\vec{\tilde{U}}_P$ is the local convection velocity around $P$, defined as:

\begin{eqnarray}
    \vec{\tilde{U}}_P = \frac{1}{S} \int_S \Vec{U}\, dS
    \label{Eq:2}
\end{eqnarray}

Since PIV measurements involve sampling the velocity field at specific discrete points, the region $S$ is defined as a rectangular area of fixed size centered around the point $P$ \cite{graftieaux2001combining}. The choice of the domain size $S$ significantly influences the accuracy of vortex boundary detection. As noted by \citet{kissing2021delaying}, a small domain size can lead to an underestimation of the vortex boundaries whereas a large domain size may introduce non-physical effects. In this paper, a domain Size $S=13$ was found to provide an optimal balance between accurately defining vortex boundary and preventing non-physical observations. The $\Gamma_2$ function is then approximated by summing over discrete spatial locations within $S$:

\begin{eqnarray}
    \Gamma_2(P) = \frac{1}{N} \sum_{S} \frac{[\vec{PM} \times (\vec{U}_M - \vec{\tilde{U}}_P)] \cdot \vec{z}}{||\vec{PM}|| \cdot ||\vec{U}_M - \vec{\tilde{U}}_P||},
    \label{Eq:3}
\end{eqnarray}

where $N$ represents the number of points $M$ inside $S$. The boundary of the vortex is determined by thresholding the $\Gamma_2$ scalar field at the value $2/\pi$. For values below $2/\pi$, the flow is locally dominated by strain, while values above indicate local dominance by rotation. The value $2/\pi$ corresponds to regions of pure shear \cite{graftieaux2001combining}. The location of the vortex center is determined by finding the local maximum of $\Gamma_2$ within the vortex boundary, reaching values between 0.9 and 1 near the vortex center, as shown in Fig. \ref{fig:2}.

Identifying the vortex boundary and center enables the calculation of the vortex ring circulation during the impact. In fact, two methods were used and compared in this study. The first method involves using Stokes' theorem to integrate the vorticity within the boundary of the detected vortex, which will be referred to as IR-$\Gamma_2$. The second method uses a least-squares best-fit on the velocity profiles, following approaches similar to those described by \citet{harris2012instability} and \citet{leweke1998cooperative}.

As the vortex ring descends after impact, the tangential velocity profile is measured along a horizontal line passing through the centers of the vortex pair, as shown by the red line in Fig. \ref{fig:2}. This profile can be described by the superposition of two Lamb-Oseen vortices (SLV), each with an azimuthal velocity profile:

\begin{eqnarray}
\begin{cases}
    u_{\phi}(r_1) = \frac{\Gamma}{2 \pi r_1} \left(1 - \exp \left(-\frac{{r_1}^2}{a^2}\right)\right), \, \text{right vortex} \\
    u_{\phi}(r_2) = \frac{-\Gamma}{2 \pi r_2} \left(1 - \exp \left(-\frac{{r_2}^2}{a^2}\right)\right), \, \text{left vortex}
    \label{Eq:4}
\end{cases}
\end{eqnarray}

where $\Gamma$ is the circulation of the vortex, $r_1$ and $r_2$ are the radial distance from the vortex center, and $a$ is a parameter characterizing the core radius of the vortex. The velocity profile along the the horizontal line can then be represented by (\ref{Eq:5}), where $d$ is vortex ring inner diameter. A least-squares best-fit was then employed with the boundary of the vortex (shaded in white in Fig. \ref{fig:3}a) to determine the circulation $\Gamma$.

\begin{widetext}
\begin{equation}
 U_y(r) = \frac{\Gamma}{2\pi (r - d/2)} \left(1 - \exp\left(-\frac{(r - d/2)^2}{a^2}\right)\right) + \left(-\frac{\Gamma}{2\pi (r + d/2)}\right) \left(1 - \exp\left(-\frac{(r + d/2)^2}{a^2}\right)\right),
    \label{Eq:5}
\end{equation}
\end{widetext}

Fig. \ref{fig:3}b shows the temporal evolution of the vortex circulation using both methods for the case $\delta = 0.90$, $We = 54$, and $Re = 3900$. The vortex characteristics were extracted from single measurements before being ensemble-averaged over multiple ones. Similar trends in the evolution of circulation are observed for both methods. Initially, circulation increases during vortex generation right after the impact, and then followed by a decrease in strength due to viscous dissipation.

However, slight variations between the methods are evident during the generation and dissipation phases. The SLV method exhibits inaccuracies, which can be attributed to several factors. First, the fitting method is more accurate when applied along the entire horizontal line of the vortex ring passing through both centers, as done in the works of \citet{harris2012instability} and \citet{leweke1998cooperative}. In this case, this is impractical since during the impact, a crater is formed creating a barrier between the two centers and complicating the velocity profile measurement across the entire line. Hence, the necessity of using the vortex boundary for the fitting model. Additionally, in this study, the vortex ring is generated by a droplet impacting on a thin liquid film, which differs from the behavior of a vortex ring moving through a deep pool, as investigated in previous studies \cite{harris2012instability, leweke1998cooperative, saha2019kinematics}. The presence of the thin film and the proximity to the wall influence the vortex dynamics and the velocity profile. These factors introduce additional complexities that may not be accurately captured by the superposition of two Lamb-Oseen vortices, suggesting that this model may not the most suitable fit for the current case.

Therefore, Stokes' theorem was preferred for calculating the circulation in this study. The uncertainty in the circulation values is showcased by the scatter plot of ensemble-averaged data obtained from multiple measurements, as shown in Fig. \ref{fig:3}b.

\section{\label{sec:4}Results and Discussion}
This section presents the experimental observations and quantitative analysis of vortex ring behavior during droplet impact on liquid films. The first part characterizes the vortex-ring dynamics that arise under varying impact conditions, leading to a regime classification based on the observed outcomes. The second part quantifies the evolution of vortex ring circulation and examines how film thickness and wall interaction influence its decay. The final part addresses the onset of azimuthal perturbations, focusing on fingering patterns and their dependence on the impact parameters.

\subsection{Vortex-ring dynamics and regime classification}
\begin{figure*}
\centering
\includegraphics[scale=0.95]{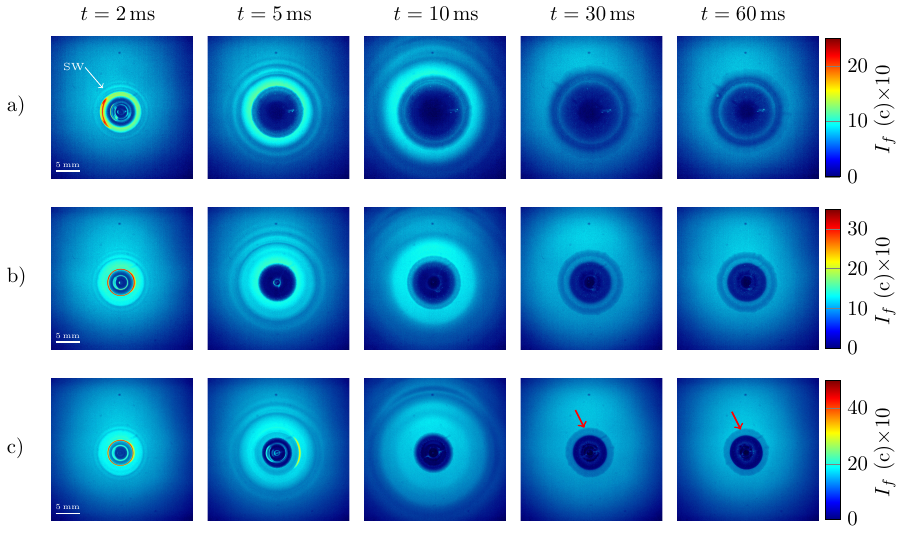}
  \caption{Time evolution of droplet impact on thin films at $Re=1400$ and $We=7$ for different film thicknesses. \textbf{a}) $\delta=0.09$, \textbf{b}) $\delta=0.29$, and \textbf{c}) $\delta=0.45$. SW marks the surface wave visible at early times, while red arrows indicated the area where floating droplet liquid causes reduced fluorescence intensity, corresponding to the features indicated with red arrows in Fig. \ref{fig:5}b. Colors represent the fluorescence signal intensity $I_f$.}
    \label{fig:4}
\end{figure*}

\begin{figure*}
\centering
\includegraphics[scale=0.85]{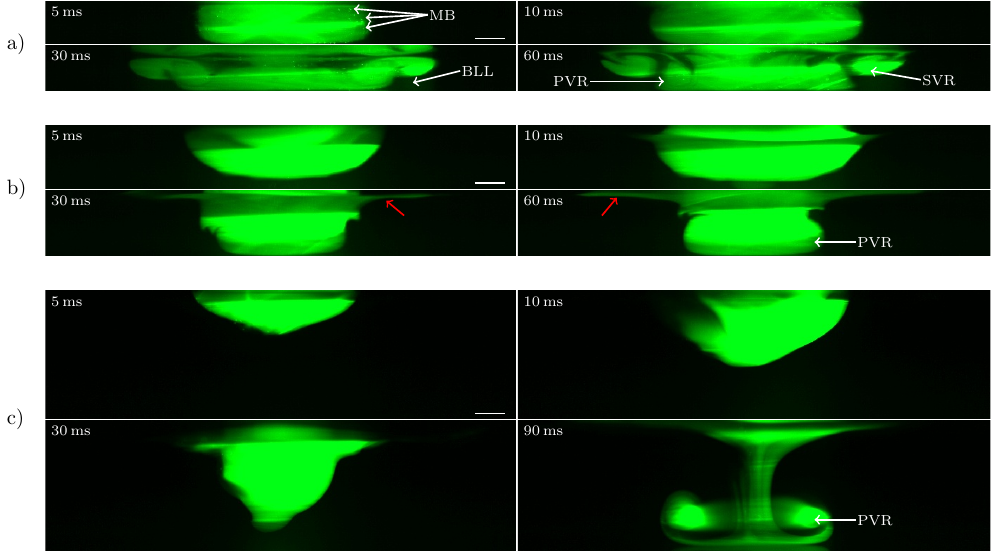}
  \caption{Side-view LIF visualizations of droplet impacts on thin films at $Re = 1400$ and $We = 7$ for different film thicknesses. \textbf{a}) $\delta=0.29$, \textbf{b}) $\delta=0.45$, and \textbf{c}) $\delta=0.90$. Images in (\textbf{a}) and (\textbf{b}) are shown at 5, 10, 30, and 60 ms after impact, while for (\textbf{c}) an additional frame at 90 ms is included. The extended time range for $\delta = 0.90$ is required because the increased film thickness delays the interaction of the vortex ring with the wall. MB marks micro-bubbles visible at early stages, BLL denotes boundary-layer lift-off, and PVR and SVR label the primary and secondary vortex rings, respectively. Red arrows in (\textbf{b}) indicate portions of droplet liquid floating on the film surface outside the vortex ring region, corresponding to the ones indicated by red arrows in Fig. \ref{fig:4}c. Scale bar is equivalent to 1 mm.}
    \label{fig:5}
\end{figure*}

To elucidate the behavior of vortex rings during droplet impact on liquid films, experiments were conducted using both bottom-view and side-view imaging. The bottom-view LIF measurements are presented as fluorescence signal intensities $I_f$ for three different dimensionless film thicknesses:  $\delta = 0.09$, $\delta = 0.29$, and $\delta = 0.45$. The side-view observations are shown as LIF visualizations for $\delta = 0.29$, $\delta = 0.45$, and $\delta = 0.90$. Side-view imaging is not feasible for $\delta=0.09$ because the film is too thin relative to the spatial resolution of the imaging system, preventing a reliable identification of the evolving impact dynamics. Exemplary images are provided in figures \ref{fig:4} and \ref{fig:5} for a Reynolds number of $Re=1400$ and a Weber number of $We=7$. All measurements cover a time span of $\SI{300}{\milli\second}$ after impact. Since the diffusion coefficient of Rhodamine B in water \cite{gendron2008diffusion} is on the order of $10^{-10}\SI{}{\square\meter/\second}$, mixing over the observed timescales is dominated by inertia-driven convection.

The evolution of fluorescence intensity during drop impact for $\delta=0.09$ is shown in figure \ref{fig:4}a. Here, decreased fluoresence intensities may either be caused by film thinning or a local increase in drop liquid which is pure water. Immediately after the initial contact at $t=\SI{2}{\milli\second}$, radially outward traveling surface waves (SW) occur inducing a local film thickening. While the impacting droplet continues to spread, an expanding crater forms with a significantly reduced liquid film thickness, resulting in a central region of low fluorescence signal $t=\SI{5}{\milli\second}$ after impact. At later times, outward-traveling surface waves have left the field of view and the crater region starts shrinking in diameter ($\SI{5}{\milli\second} \le t \le \SI{60}{\milli\second}$). At these time instances, concentric circular intensity pattern outside the crater region can be attributed to concentration variations with darker regions corresponding to higher concentration droplet liquid.

For the given Reynolds and Weber number regime rotational symmetry is maintained also for non-dimensional film thicknesses $\delta=0.29$ and $\delta=0.45$ as indicated in Figs. \ref{fig:4}b and c. For increasing film thicknesses, the crater region assumes smaller values of maximum radius. This reduction is a consequence of earlier momentum redirection toward the radial direction in thinner films. In fact, in thinner films, wall interactions occurs sooner during the impact, causing the remaining kinetic energy to be redirected radially at an earlier stage of the impact. Hence, the larger surface of mixing.

Complementary side-view visualizations are given in Figures \ref{fig:5}a-c. A mentioned before, for these side view measurements, the drops are dyed, shown here in green colour. Figure \ref{fig:5}a corresponds to Fig. \ref{fig:4}b for $\delta = 0.29$ gaining further details on the vortex ring formation and evolution upon droplet impact. 
The sequence of the images shows the droplet penetrating the liquid film and forming a vortex ring that approaches the wall. Between $t=\SI{5}{\milli\second}$ and $t=\SI{10}{\milli\second}$, a vortex ring has formed that spreads radially due to interaction with the wall as already discussed in Section \ref{sec:1}. The expanding vortex ring triggers boundary layer lift-off (BLL) from which a secondary vortex ring forms with opposite vorticity. The emergence of such a secondary vortex is visible in Fig. \ref{fig:5}a at $t=\SI{60}{\milli\second}$. From these side-view visualizations, it becomes evident that the concentric circular concentration patterns seen in the bottom-view images originate from the primary (PVR) and secondary (SVR) vortex rings, each transporting different proportions of droplet and film liquid. In particular, the primary vortex ring appears as a darker circular region because it predominantly carries droplet liquid.

For $\delta=0.45$, the side-view observations (Fig. \ref{fig:5}b) reveal that no boundary layer lift-off takes place, such that only one primary vortex is formed that drives the mixing process. Additionally, portions of the droplet liquid are seen floating on the surface outside the vortex ring region at both $t=\SI{30}{\milli\second}$ and $t=\SI{60}{\milli\second}$, explaining regions of reduced brightness in the bottom-view images. These regions are indicated by red arrows in both the bottom-view and side-view visualizations. For $\delta=0,90$, which represents a thick film \cite{geppert2019experimental}, the side-view images in Fig. \ref{fig:5}c illustrate how the droplet penetrates the film, forming a descending vortex ring that moves toward the wall. Similarly to $\delta=0.45$, the vortex ring expands near the wall, but no boundary layer separation is observed. Consequently, in thicker films, the influence of the wall on the vortex ring dynamics is less pronounced, allowing the primary vortex ring to retain its coherence without generating additional vortical structures.

An additional phenomenon observed in low Weber number impacts is the presence of micro-bubbles (MB) immediately after impact, visible in Figs \ref{fig:5}a. Similar effects have been reported by Thoroddsen \textit{et al.}\cite{thoroddsen2003air} during droplet impact on a pool at $We<22$. They attributed this to the formation of a thick air sheet trapped between the droplet and the liquid surface due to the large deformations of the free surface prior to contact. 

\begin{figure*}
\centering
\includegraphics[scale=0.95]{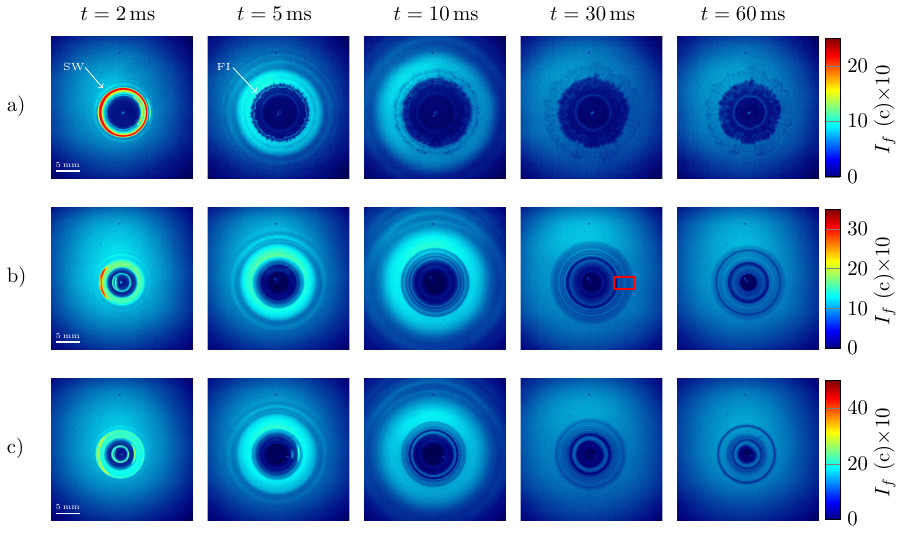}
  \caption{Time evolution of droplet impact on thin films at $Re=1900$ and $We=13$ for different film thicknesses. \textbf{a}) $\delta=0.09$, \textbf{b}) $\delta=0.29$, and \textbf{c}) $\delta=0.45$. SW marks the surface wave visible at early times, while FI indicates the finger-shaped instabilities. The red box in (\textbf{b}) highlights the region corresponding to the boxed area in the side-view visualization of Fig. \ref{fig:7}a. Colors represent the fluorescence signal intensity $I_f$.}
    \label{fig:6}
\end{figure*}

\begin{figure*}
\centering
\includegraphics[scale=0.85]{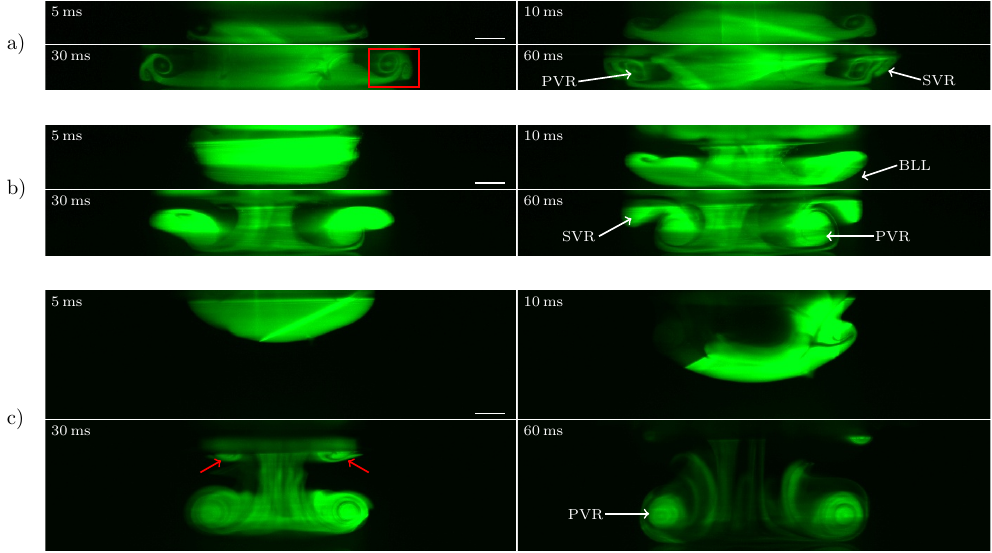}
  \caption{Side-view LIF visualizations of droplet impacts on thin films at $Re = 1900$ and $We = 13$ for different film thicknesses. \textbf{a}) $\delta=0.29$, \textbf{b}) $\delta=0.45$, and \textbf{c}) $\delta=0.90$. The time series corresponds to 5, 10, 30, and 60 ms after impact. PVR and SVR denote the primary and secondary vortex rings, respectively, while BLL marks the boundary-layer lift-off observed. The red box in (\textbf{a}) highlights the region that corresponds to the boxed area in the bottom-view image of Fig. \ref{fig:6}b. Red arrows in (\textbf{c}) indicate the near-surface vortex rings formed during the crater receding phase. Scale bar is equivalent to 1 mm.}
    \label{fig:7}
\end{figure*}

\begin{figure*}
\centering
\includegraphics[scale=0.95]{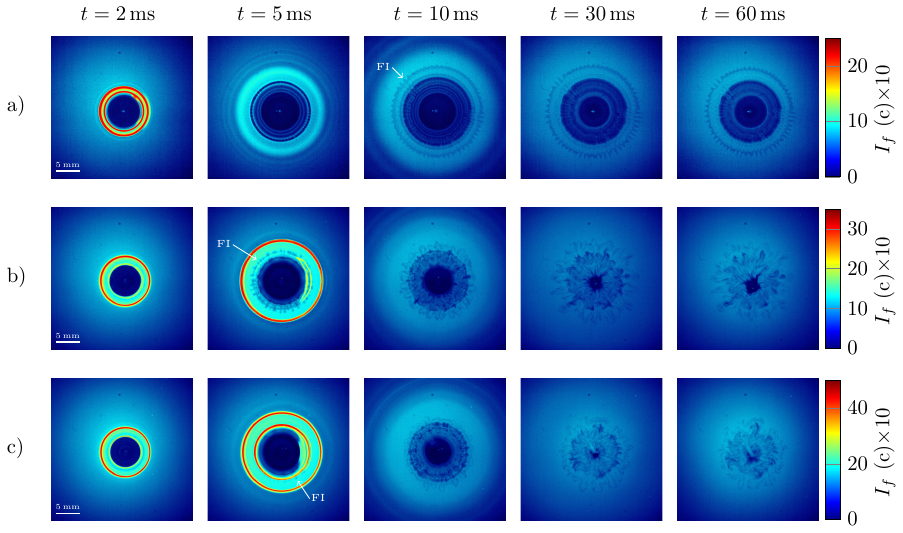}
  \caption{Time evolution of droplet impact on thin films at $Re=3900$ and $We=54$ for different film thicknesses. \textbf{a}) $\delta=0.09$, \textbf{b}) $\delta=0.29$, and \textbf{c}) $\delta=0.45$. SW marks the surface wave visible at early times, while FI indicates the finger-shaped instabilities. Colors represent the fluorescence signal intensity $I_f$.}
    \label{fig:8}
\end{figure*}

\begin{figure*}
\centering
\includegraphics[scale=0.85]{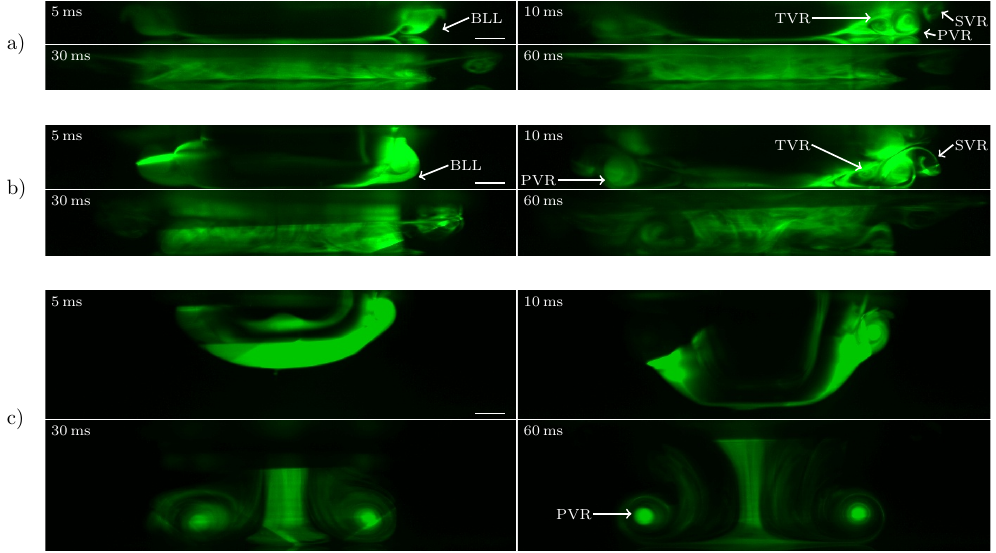}
  \caption{Side-view LIF visulizations of droplet impacts on thin films at $Re = 3900$ and $We = 54$ for for different film thicknesses. \textbf{a}) $\delta=0.29$, \textbf{b}) $\delta=0.45$, and \textbf{c}) $\delta=0.90$. The time series corresponds to 5, 10, 30, and 60 ms after impact. PVR, SVR and TVR denote the primary, secondary and tertiary vortex rings, respectively, while BLL marks the boundary-layer lift-off observed. Scale bar is equivalent to 1 mm.}
    \label{fig:9}
\end{figure*}

\begin{figure*}
\centering
\includegraphics{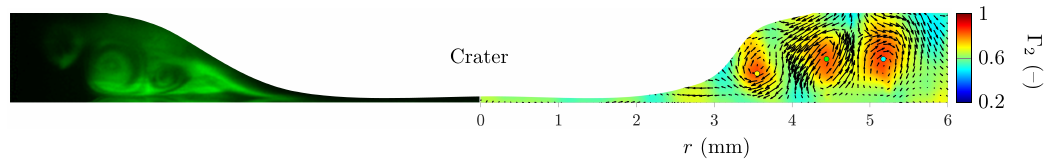}
  \caption{(Left) Visualization of the vortex ring during droplet impact on liquid film using Laser-Induced Fluorescence. (Right) Velocity vector field overlaid with $\Gamma_2$ scalar field exctracted from PIV measurements. The three point indicates the centers of the vortices. The case corresponds to $\delta = 0.29$, $We = 54$, $Re = 3900$ at $t=\SI{10}{\milli\second}$.}
    \label{fig:10}
\end{figure*}

\begin{figure}[htbp]
    \centering
    \includegraphics{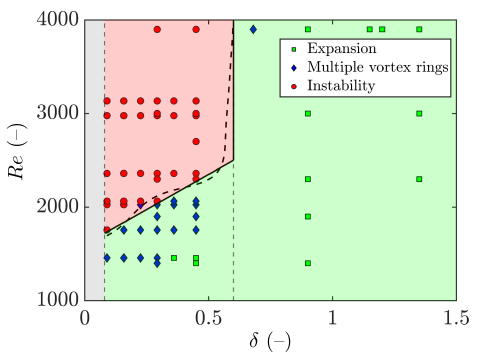}
       \centering
    \caption{Regime map of vortex ring outcomes during droplet impact on liquid films. Dashed black line: regime-boundary derived from support vector machine (SVM) classification \citet{cortes1995support}. Solid black line: simplified representation of this boundary based on Eq. \ref{Eq:6} and on the transition at $\delta=0.6$, which marks the shift from thin films to thick film regimes. The green region includes "Expansion" and "Multiple vortex rings", while the red region indicates vortex ring fingering/decomposition. Dashed grey vertical lines represent the thin film regimes boundary $0.08< \delta <0.60$. Film thicknesses below $\delta = 0.08$ are shaded in gray and were not considered in this study.}
    \label{fig:11}
\end{figure}

Figures \ref{fig:6} and \ref{fig:7} present the bottom-view and side-view images, respectively, for an increased Reynolds and Weber number of $Re=1900$ and $We=13$. For $\delta=0.09$, shown in Fig. \ref{fig:6}a, finger-shaped instabilities (FI) are observed along the perimeter of the primary vortex ring during the spreading phase at $t=\SI{5}{\milli\second}$. Wavy instabilities also appear on the secondary vortex ring at $t=\SI{10}{\milli\second}$, similar to observations made by Walker and others \cite{walker1987impact, cerra1983experimental,mishra2021instability} where vortex rings interact with walls during high-energy impacts. 
In the case of $\delta = 0.29$, shown in Fig. \ref{fig:6}b, multiple concentric rings begin to emerge at $t = \SI{10}{\milli\second}$. This pattern is attributed to the formation of the secondary vortex ring, as confirmed by the side-view images in Fig. \ref{fig:7}a. The regions highlighted in the red boxes show how the spiral motion of of the primary and secondary vortex rings entrains film and droplet liquid in different proportions, creating concentration variations that appear in the bottom view as successive concentric rings. This phenomenon is also observed for increased film thickness of $\delta=0.45$ (Fig. \ref{fig:6}c, and \ref{fig:7}b). However, a notable difference lies in the core size of the vortex rings, which increases with film thickness. At $\delta=0.90$, as shown in Fig. \ref{fig:7}c, no boundary layer separation is observed. However, an additional aspect becomes visible after $t=\SI{10}{\milli\second}$. The primary vortex ring has not yet reached the wall when another vortex ring appears near the surface of the liquid film, indicated by red arrows. This ring is generated during the receding phase of the crater and is not a result of the primary vortex ring interacting with the wall. The upward movement of the droplet liquid as the crater recedes induces vorticity, creating a vortex ring positioned just beneath the film's surface.

\begin{figure*}
\centering
\includegraphics{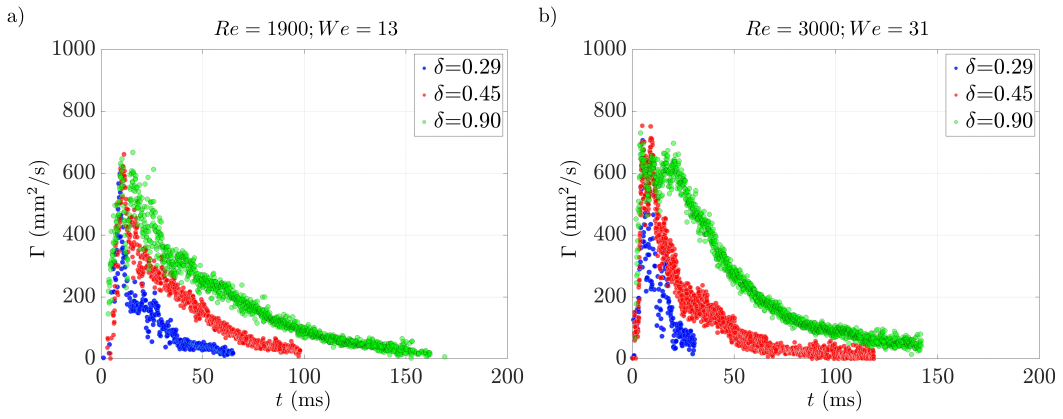}
  \caption{Temporal evolution of the primary vortex ring circulation for different film thicknesses: $\delta = 0.29$ (blue), $\delta = 0.45$ (red), and $\delta = 0.90$ (green). \textbf{a)} $Re = 1,900$, $We = 13$. \textbf{b)} $Re = 3,000$, $We = 31$.}
    \label{fig:12}
\end{figure*}

Figure \ref{fig:8} and \ref{fig:9} display the images of drop impacts from a bottom and side perspective for the highest investigated Reynolds and Weber number of $Re=3900$ and $We=54$. For $\delta=0.09$, as depicted in Fig. \ref{fig:8}a, finger-shaped instabilities similar to those at $Re=1900$ can be observed. However, the wavelength of these instabilities is shorter resulting in a higher number of fingers. This pattern resembles the fingering pattern observed during droplet impact onto dry surfaces, as reported by \citet{marmanis1996scaling} and \citet{thoroddsen1998evolution}.


At higher film thicknesses of $\delta=0.29$ (see Fig. \ref{fig:8}b), pronounced azimuthal instabilities are observed. These instabilities result in chaotic mixing patterns. Furthermore, side-view images for the same $\delta$, $Re$ and $We$ in Fig. \ref{fig:9}a reveal the formation of three distinct vortex rings. At $t=\SI{10}{\milli\second}$, these three rings are clearly visible both in the LIF visualizations and in the $\Gamma_2$ scalar field shown in Fig. \ref{fig:10}. At $t=\SI{5}{\milli\second}$, the tertiary vortex ring begins to appear as a vortical structure generated by shear flow along the wall, consistent with a Kelvin–Helmholtz–type mechanism. For clarity, the primary, secondary, and tertiary labels follow the chronological order in which the vortex rings are identified, based on the time-resolved PIV measurements and the vortex identification method. Once formed, the interaction between these vortex rings lead to their three-dimensional breakdown, consistent with the elliptical (TWMS) instability\cite{tsai1976stability,moore1975instability} described in Section \ref{sec:1}. This decay contributes to enhanced mixing in the liquid film as documented in our previous work \cite{ennayar2023lif}. For $\delta=0.45$, presented in Figs. \ref{fig:8}c and \ref{fig:9}b, similar vortex pattern is observed, with multiple vortex ring leading to chaotic mixing. The major difference is the reduced area of mixing compared to $\delta=0.29$, which is attributed to the increased film thickness.

In the thick film case of $\delta=0.90$, shown in Fig. \ref{fig:9}c, the primary vortex ring reaches the wall at approximately $t=\SI{25}{\milli\second}$. No boundary layer separation or formation of a secondary vortex ring is observed at $t=\SI{30}{\milli\second}$ and $t=\SI{60}{\milli\second}$. The primary vortex ring retains its coherence due to the absence of additional vortical structures.

From the parameter study conducted, three distinct vortex ring outcomes during droplet impact on liquid films were identified: the expansion of the primary vortex ring upon collision with the wall (referred to as "Expansion"), the formation of secondary and sometimes tertiary vortex rings during the spreading of the primary vortex ring (referred to as "Multiple vortex rings"), and the onset of vortex ring instabilities with circumferential fingering pattern or decomposition (referred to as "Instability"), which can manifest as either long-wavelength or short-wavelength perturbations. Based on the parameter study a regime map was established as plotted in Fig. \ref{fig:11}, representing the influence of Reynolds number $Re$ and the dimensionless film thickness $\delta$ on the observed outcomes. The transition from expansion to multiple vortex rings and eventually to fingering pattern and chaotic vortex disintegration occurs at lower $Re$ for thinner films compared to thicker ones. Notably, in the range where $\delta > 0.60$, corresponding to the thick film regime, no vortex ring instability was observed across the entire range of impact conditions within the deposition regime.

These outcomes were also classified into two main regions. The vortex ring region (green) encompasses both the expansion and multiple vortex ring outcomes, while the fingering/chaotic region (red) indicates the occurrence of vortex ring instabilities. A Support Vector Machine (SVM), introduced by \citet{cortes1995support}, was employed to generate the decision boundary that differentiate between the two regions. SVM is a widely recognized machine learning method commonly applied to solve classification problems \cite{awad2015efficient,stumpf2022imaging}. The SVM classification produced a non-linear boundary capturing the combined effects of $Re$ and $\delta$. The non-linear boundary, derived from the SVM classification, reflects the influence of $Re$ and $\delta$. However, to enhance interpretability, Fig. \ref{fig:11} present a simplified representation of this boundary. A significant factor in this simplification is the transition at $\delta=0.60$, marking the shift from thin film to thick film regimes. It is evident that beyond this threshold, no vortex ring instability occurs. Additionally, the use of a linear boundary for the thin film regime suggests a simpler relationship between $Re$ and $\delta$, which can be approximated by Eq. (\ref{Eq:6}):

\begin{eqnarray}
    Re = 1600 + 1500 \cdot \delta
    \label{Eq:6}
\end{eqnarray}

Film thicknesses below $\delta=0.08$ were not considered in this study and are shaded in gray. It should be noted that the boundary, while consistent with the available data, remains approximate and requires further experimental confirmation, particularly through measurements conducted in close proximity to the predicted transition region.

\begin{figure}[htbp]
    \centering
    \includegraphics{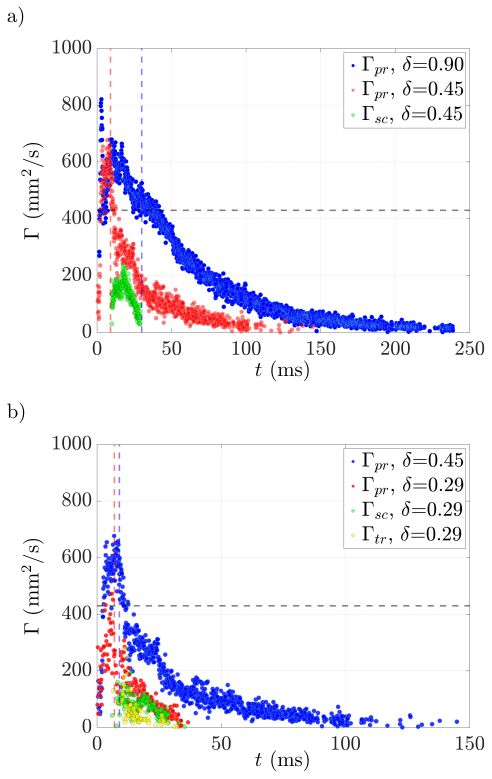}
       \centering
    \caption{Temporal evolution of the vortex ring circulation for the case $Re=3900$ and $We=54$. \textbf{a)} Primary vortex ring for $\delta=0.90$ (blue), primary vortex ring for $\delta=0.45$ (red) and secondary vortex ring for $\delta=0.45$ (green). \textbf{b)} Primary vortex ring for $\delta=0.45$ (blue), primary vortex ring for $\delta=0.29$ (red), secondary vortex ring for $\delta=0.29$ (green) and tertiary vortex ring for $\delta=0.29$ (yellow).}
    \label{fig:13}
\end{figure}

\subsection{Evolution of circulation decay in vortex rings}
To gain quantitative insights into the different outcomes observed during droplet impact on liquid films, the circulation decay of the vortex rings was investigated. If a vortex ring collides with a wall in a semi-infinite domaine, a relevant dimensionless number is the circulation Reynolds number \cite{harris2012instability,mishra2021instability} $Re_\Gamma$. The present study aims to explore the extent to which film thickness influences the instability of vortex rings generated by droplet impact, as well as its temporal evolution under various impact conditions.

\begin{figure*}
\centering
\includegraphics{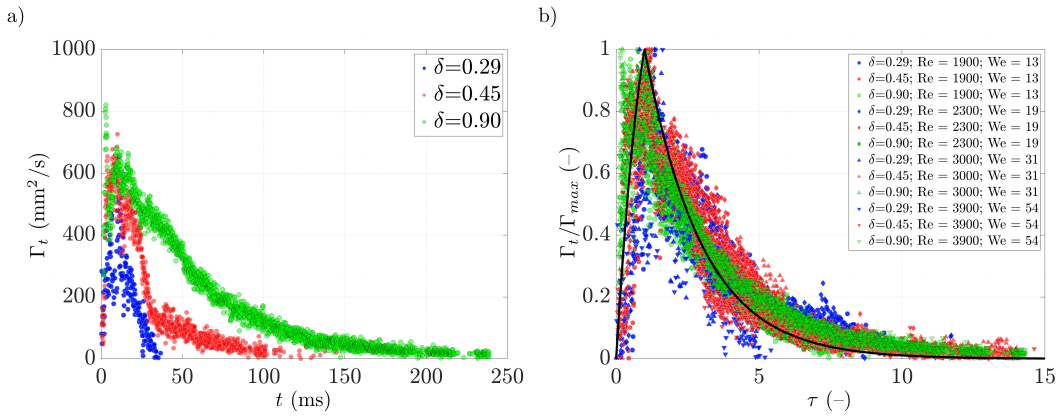}
  \caption{\textbf{a)} Time evolution of the total ciruculationfor the case $Re=3900$ and $We=54$ for different film thicknesses: $\delta = 0.29$ (blue), $\delta = 0.45$ (red), and $\delta = 0.90$ (green). \textbf{b)} Time evolution of the dimensionless total circulation plotted against the dimensionless time defined by Eq. \ref{Eq:13} for all investigated cases. The black curve corresponds to the empirical fit given by Eq. \ref{Eq:17}.}
    \label{fig:14}
\end{figure*}

The circulation of vortex rings was calculated from PIV measurements using Stokes' theorem, as described in Section \ref{sec:3}. Figure \ref{fig:12} shows the temporal evolution of the primary vortex ring circulation under two different impact conditions for three different film thickness, $\delta=0.29$ (blue), $\delta=0.45$ (red) and $\delta=0.90$ (green). Fig. \ref{fig:12}a corresponds to $Re=1900$ and $We=13$, while Fig. \ref{fig:12}b corresponds to $Re=3000$ and $We=31$. For both sets of impact conditions, all curves exhibit an initial sharp rise in circulation shortly after droplet impact, reaching a peak before gradually decaying over time. This initial peak corresponds to the generation of the primary vortex ring resulting from the droplet impact. Additionally, the influence of film thickness on the circulation evolution is clearly observed. Focusing on the case of $Re=1900$ (Fig. \ref{fig:12}a), the impact energy is similar for all three thicknesses, as indicated by the comparable peak circulation values. For the thick film ($\delta=0.90$), a gradual decay of circulation is observed, primarily due to viscous dissipation. In the case of $\delta=0.45$, a slightly steeper decay around $t=\SI{12}{\milli\second}$, corresponding to the time when the primary vortex ring collides with the wall. This interaction results in additional energy dissipation due to boundary layer effects and the formation of a secondary vortex ring, explaining the increase decay compared to the thicker film. Further decreasing the film thickness to $\delta=0.29$, an even more pronounced decay in the strength of the primary vortex ring is observed. The thinner film leads to earlier wall interaction, resulting in the generation of a secondary vortex ring and even more stronger viscous effect, which contribute to the accelerated decay of circulation.

\begin{figure}[htbp]
    \centering
    \includegraphics{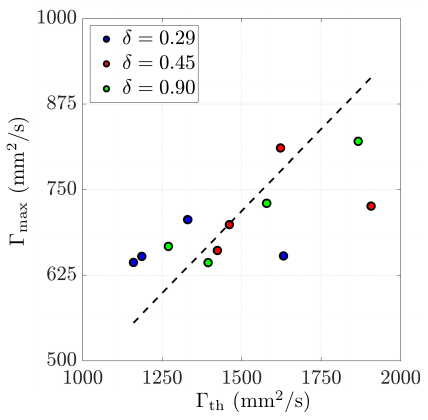}
       \centering
    \caption{Maximum measured total circulation $\Gamma_{\text{max}}$ plotted against the energy-based model circulation $\Gamma_{\text{th}}$ with the dashed line representing the slope $\sqrt{\xi}$. Colors denote dimensionless film thicknesses.}
    \label{fig:15}
\end{figure}

When comparing these observations with a higher Reynolds number case $Re=3000$ shown in Fig. \ref{fig:12}b, the gap between the thinner films and the thicker one becomes more significant. This is attributed to instabilities developing in the primary vortex ring, leading to its rapid decay. $\delta=0.90$ continues to exhibit a slow decay due to viscosity , with minimal influence from the wall interactions. In contrast, for $\delta=0.29$, the structure of the primary vortex ring breaks down entirely into three-dimensional chaotic mixing around $t=\SI{30}{\milli\second}$ and can no longer be detected.

Figure \ref{fig:13} further illustrates the temporal variation of circulation for the formed vortex rings after wall interaction , alongside the primary vortex ring's circulation for $Re=3900$ and $We=54$. Under these conditions, instability occurs for thin films, whereas the primary vortex ring remains coherent in thicker films. Fig. \ref{fig:13}a compares the temporal evolution of primary vortex ring circulation for $\delta=0.90$ (blue) and $\delta=0.45$ (red), as well as the circulation of the secondary vortex ring for $\delta=0.45$ (green). Shortly after droplet impact, both primary vortex rings exhibit an increase in circulation. Around $t=\SI{8}{\milli\second}$, indicated by the red dashed line, a sharp decrease in the circulation of the primary vortex ring for $\delta=0.45$ is observed, accompanied by an increase in the circulation of the secondary vortex ring. This behavior corresponds to the formation of the secondary vortex ring due to wall interaction, as confirmed by side-view visualizations. The secondary vortex ring gains strength and subsequently starts decaying due to instabilities and viscous dissipation until it breaks down. The dashed blue line in Fig. \ref{fig:13}a indicates the time at which the primary vortex ring for $\delta=0.90$ reaches the wall. However, minimal effect is observed on its circulation evolution, suggesting that wall interaction is less significant for thicker films. 

Additionally, a horizontal dashed black line in Fig. \ref{fig:13}a represents the critical $Re_\Gamma$ value of 430 identified by \citet{harris2012instability}. In their investigation of a vortex pair approaching a solid boundary in a homogeneous fluid domain, rather than a thin liquid film, they showed that below this critical value no secondary vortical structures are generated during wall collision. For $\delta=0.45$ in the present work, the primary vortex ring reaches the wall with a strength equivalent to $\Gamma_{pr}=\SI{665}{\milli\meter}^2/\SI{}{\second}$, corresponding to $Re_\Gamma=665$, exceeding the critical threshold and agreeing with the observed formation of secondary vortical structures. In the thicker film case $\delta=0.90$, the strength of the primary vortex ring is close to the critical value, still no secondary vortex ring was observed. An additional film thickness of $\delta=0.68$ was investigated, which falls within the thick film regime. This case exhibited the formation of a secondary vortex ring (see Fig. \ref{fig:11}) after wall collision but did not display any instabilities. The evolution of circulation for $\delta=0.90$ can thus serve as a benchmark for other film thicknesses before wall collision. The thinner the film, the earlier the wall interaction occurs, and if the vortex strength at that times exceeds the critical threshold, secondary vortex ring is formed. Another aspect is observed for $\delta=0.29$ in Fig. \ref{fig:13}b. As previously discussed, a third vortex ring is created due to the Kelvin-Helmholtz instability. The creation of this third vortex ring (yellow curve) leads to a decrease in the peak circulation of the primary vortex ring (red curve). Moreover, the circulation of the latter exceeds the critical threshold, leading to the formation of a secondary vortex ring (green curve). The three vortex rings decay more rapidly compared to the primary vortex ring of $\delta=0.45$ (blue curve) and ultimately break down into turbulent three-dimensional mixing.

Next, a empirical fit was developed to predict the time evolution of the total circulation of the vortex rings generated in each case. For each specific case with a unique $\delta$, $Re$ and $We$, the circulations of all generated vortex rings were extracted from the PIV data and summed to obtain the total circulation $\Gamma_t=\sum \Gamma$. Fig. \ref{fig:14}a presents the total circulation for each thickness under the impact condition of $Re=3900$ and $We=54$.A striking observation is that, even after summing the overall circulation of the vortex rings for $\delta=0.29$ and $\delta=0.45$, a sharp decrease in circulation is still observed. This decrease is explained by the energy dissipation into three-dimensional turbulent structures as seen in the visualizations, leading to enhanced mixing \cite{ennayar2023lif}.

To interpret the magnitude of the total measured circulation, a kinematic energy balance was considered. The droplet possesses an initial kinetic energy
\begin{eqnarray}
    E_d = \frac{\pi}{12} \rho D^3 U^2.
    \label{Eq:7}
\end{eqnarray}

The kinetic energy of a thin-core, axisymmetric vortex ring is given by \cite{roberts1970dynamics,donnelly1991quantized}
\begin{eqnarray}
    E_d = \frac{1}{4} \rho \Gamma^2 d \left[ \text{ln}(\frac{4d}{a}) - \alpha \right],
    \label{Eq:8}
\end{eqnarray}
with $\alpha=7/4$. The equation is derived under the assumptions of an incompressible, inviscid fluid, and a circular ring of slender core ($a\ll d/2$)\cite{sullivan2008dynamics}. Because only part of the droplet kinetic energy is transferred into the vortex structure , the remaining energy being dissipated through interfacial deformation and shear, an efficiency factor $\xi$ as introduced:
\begin{eqnarray}
    E_v = \xi E_d.
    \label{Eq:9}
\end{eqnarray}

Solving the balance for $\Gamma$ yields
\begin{eqnarray}
    \Gamma_{\text{max}} = \sqrt{\xi}\Gamma_{\text{th}}=\sqrt{\xi}\cdot UD^{3/2}\sqrt{\frac{\pi}{3d\left[ \text{ln}(\frac{4d}{a}) - \alpha\right]}}.
    \label{Eq:10}
\end{eqnarray}

The parameter $\xi$ was determined empirically by comparing $\Gamma_{\text{th}}$ with the maximum measured total circulation $\Gamma_{\text{max}}$ extracted from PIV measurements. Across all investigated conditions, $\xi=0.222$ provided the best agreement as illustrated in Fig. \ref{fig:15}. Moreover, the choice of the core radius $a$ depended on the film thickness. For thin films ($\delta<0.6$), $a=h/2$ was used, as the inner core diameter is limited by the film thickness. For thicker films, $a$ was taken as the measured inner core radius of the vortex ring.

Furthermore, a classical scale of the viscous diffusion of vorticity across a layer of thickness $h$ is 
\begin{eqnarray}
    t_{\nu} \sim \frac{h^2}{\nu}.
    \label{Eq:11}
\end{eqnarray}

However, the decay rate observed in the experiments depends not only on viscous diffusion but also by the combined effect of vortex strength and geometric confinement. A stronger vortex impinges on the wall more rapidly when the film is thin, leading to enhanced dissipation. These effects enter through the circulation-based Reynolds number $Re_{\Gamma_\text{max}}= \Gamma_{\text{max}}/\nu$ and the dimensionless film thickness $\delta$. Accounting for both these effects leads to a corrected characteristic time scale
\begin{eqnarray}
    t_{c} \sim \frac{t_{\nu}}{Re_{\Gamma_\text{max}}\delta} = \frac{h^2}{\nu}\frac{\nu}{\Gamma_\text{max}}\frac{1}{\delta}=\frac{D^2\delta}{\Gamma_\text{max}}.
    \label{Eq:12}
\end{eqnarray}

This time scale decreases for stronger vortex rings and for thinner films, consistent with the accelerated circulation decay observed in these cases, where the kinetic energy of the vortical structures is dissipated predominantly through viscous diffusion in thick films and additionally through the breakdown into three-dimensional turbulent structures in thin films.

The temporal evolution of the total circulation was analyzed in dimensionless form $\Gamma_t / \Gamma_{\text{max}}$. The dimensionless time is defined as:
\begin{eqnarray}
    \tau = \frac{t}{t_c} = t \left( \Gamma_{\text{max}} / (D^2 \delta) \right).
    \label{Eq:13}
\end{eqnarray}
Figure \ref{fig:14}b shows that the temporal circulation for all cases displays similar behavior during the vortex ring generation and decay phases. On average, the maximum circulation is reached at $\tau_{\text{max}}=1$. Moreover, The progression of the circulation can be decomposed into two parts, an inertia-driven phase, where the primary vortex ring is generated until reaching its maximum strength, and a decay phase, where the circulation decreases due to various factors discussed earlier. To model the behavior of the total circulation over time, an empirical fit using sigmoid functions was employed. The sigmoid function \cite{verhulst1838notice}, also known as the logistic function, is an S-shaped curve defined as:

\begin{eqnarray}
    S(\tau) = \frac{1}{1 + \exp(-k(\tau - \tau_{\text{max}}))},
    \label{Eq:14}
\end{eqnarray}

where $k$ is a parameter controlling the steepness of the transition. The sigmoid function transitions between the values $0$ and $1$, allowing the modeling of the switch between the inertia-driven and decay phases. The dimensionless total circulation is then expressed by Eq. (\ref{Eq:15}), where $A(\tau)$ represents the inertia-driven phase and $B(\tau)$ represents the decay phase.

\begin{eqnarray}
    \frac{\Gamma_t}{\Gamma_{\text{max}}} = A(\tau) [1 - S(\tau)] + B(\tau) S(\tau).
    \label{Eq:15}
\end{eqnarray}

The sigmoid function modulates the contribution of each phase depending on $\tau$. For $\tau \ll \tau_{\text{max}}$, the inertia-driven part dominates, while for $\tau \gg \tau_{max}$, the decay part becomes significant. Inspired by the work of Roisman \textit{et al.}\cite{roisman2008propagation}, who showed how the temporal evolution of the crater diameter follows a parabolic curve, the inertia-driven phase is then modeled by Eq. (\ref{Eq:16}), where $b_1$ and $b_2$ are fitting parameters. The decay phase is expressed as an exponential decay:

\begin{eqnarray}
\begin{cases}
    A(\tau) = b_1 \tau^2 + b_2 \tau, \\
    B(\tau) = \exp(-c (\tau - \tau_{\text{max}})),
    \label{Eq:16}
\end{cases}
\end{eqnarray}

where $c$ is a decay constant. By combining these expressions, the full empirical model becomes as follows:

\begin{widetext}
\begin{equation}
 \frac{\Gamma_t}{\Gamma_{\text{max}}} = \left( (\frac{1 - \tau_{\text{max}} b}{\tau_{\text{max}}^2}) \tau^2 + b \tau \right) \left[1 - \frac{1}{1 + \exp(-k(\tau - \tau_{\text{max}}))}\right] + \exp\left(-c(\tau - \tau_{\text{max}})\right) \frac{1}{1 + \exp(-k(\tau - \tau_{\text{max}}))}.
    \label{Eq:17}
\end{equation}
\end{widetext}

The parameter obtained from fitting data are $b=1.5$, $c=0.5$ and $k=1281$. As can be seen in Fig. \ref{fig:14}b, the black curve, representing the empirical fit, aligns well with the temporal evolution of the total circulation. Additionally, an interesting observation is that, for both cases $\delta=0.29$ and $\delta=0.45$ at $Re=3900$, the circulation values deviates slightly, reaching lower values. This deviation is attributed to the instabilities leading to chaotic mixing, which enhances energy dissipation.

\subsection{Azimuthal perturbations and fingering mechanisms}
The instabilities occurring in vortex rings during droplet impact on thin liquid films were investigated, focusing on the number of fingers formed and the influence of the studied parameters on this phenomenon. In the standard case of a vortex ring colliding with a wall in a semi-infinite domaine, instabilities typically initiate in the secondary vortex ring \cite{walker1987impact,cerra1983experimental}. \citet{walker1987impact} observed that primary vortex rings remain initially stable prior to collision with the wall and do not exhibit waviness until instabilities manifest first in the secondary vortex ring. Further observation by \citet{cerra1983experimental} indicated that as the secondary vortex ring forms, it begins to orbit around the primary vortex ring towards its inner perimeter. This latter leads to a compression of the diameter of the secondary vortex ring, causing instabilities as explained by \citet{saffman1979vortex}.

In the context of droplet impact on thin liquid films, the dynamics differ significantly. A notable observation is that the secondary vortex ring, once it starts orbiting, does not reach the interior of the primary vortex ring. This limitation is imposed by the finite thickness of the film. Consequently, the mechanism leading to instabilities in this case is hypothesized to differ from that in a semi-infinite domaine. It is suggested that the compression in the diameter of the vortex rings remains the primary cause of the instabilities, however, the mechanism inducing this compression is attributed to the receding motion of the crater. As the crater recedes, it entrains the vortex rings, leading to a reduction in their diameter and triggering instabilities in both primary and secondary vortex rings. This strong compression effect is more pronounced in thinner films, where the influence of the crater dynamics is intensified.

\begin{figure}[htbp]
    \centering
    \includegraphics{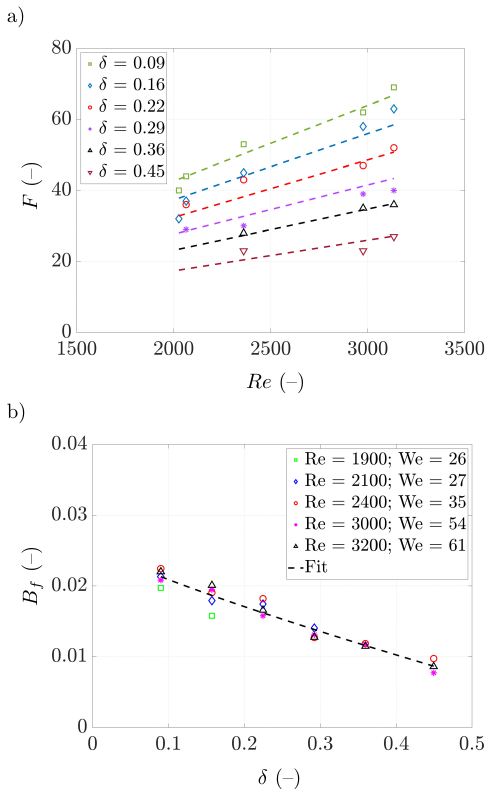}
       \centering
    \caption{\textbf{a)} Correlation between the number of fingers $F$ and the Reynolds number for different film thicknesses. \textbf{b)} Coefficient $B_f$ plotted as a function of the dimensionless film thickness $\delta$ for different Reynolds numbers. The dashed line corresponds to the fit given by Eq. \ref{Eq:19}.}
    \label{fig:16}
\end{figure}

Next, the number of fingers resulting from the instabilities was analyzed, as it is directly connected to the wavelength of the wavy perturbations on the vortex rings. Mishra \textit{et al.}\cite{mishra2021instability} stated that the characteristic wavelength of the instabilities is related to the core radius of the vortex ring defined by $a$ in the section \ref{sec:3}. Thinner rings possess smaller core radii, leading to a shorter wavelength, and consequently, a greater number of fingers. Figure \ref{fig:16}a illustrates the correlation between the number of fingers $F$ and the Reynolds number for different film thicknesses. Notably, the number of fingers increases as the film thickness decreases. This observation aligns well with the earlier statement regarding characteristic wavelength \cite{mishra2021instability}, since impacts on thinner films produced vortex rings with smaller core sized compared to those generated in thicker films, as impacts on thinner films produce vortex rings with smaller core sizes compared to those generated in thicker films, as discussed previously. Additionally, the number of fingers increases with $Re$ in a linear manner, which can be expressed as: 

\begin{eqnarray}
    F = B_f \cdot Re,
    \label{Eq:18}
\end{eqnarray}

where $B_f$ is a coefficient dependent on the dimensionless film thickness $\delta$. The relationship between between $B_f$ and $\delta$ was determined empirically and is written as follow:

\begin{eqnarray}
    B_f = 0.02573 - 0.0334 \cdot \delta^{0.8398}.
    \label{Eq:19}
\end{eqnarray}

In Figure \ref{fig:16}b, the coefficient $B_f$ is plotted against $\delta$ for different Reynolds numbers, which shows a strong correlation between the values of $B_f$ obtained from Eq. (\ref{Eq:18}) and those predicted by Eq. (\ref{Eq:19}) across the range of parameter studied. 

\section{\label{sec:5}Conclusion}
In this study, for the first time the dynamics of vortex rings during droplet impact on liquid films were investigated experimentally using Particle Image Velocimetry (PIV) and Laser-Induced Fluorescence (LIF). By varying the Reynolds number $Re$, Weber number $We$ and dimensionless film thickness $\delta$, the influence of these parameters on vortex ring behavior was examined across both thin and thick film regimes ranging from $\delta=0.09$ up to $\delta=1.35$.

Three distinct outcomes of vortex ring were identified: the expansion of the primary vortex ring upon collision with the wall, the formation of multiple vortex rings due to boundary layer separation, and the onset of instabilities manifesting as azimuthal finger-shaped perturbations. Notably, it was observed that the transition from expansion to instability occurs at lower $Re$ for thinner films, whereas no instabilities were detected for film thicknesses exceeding $\delta=0.6$. This indicates that the additional boundary imposed by the wall enhances and triggers destabilization. To visualize/capture the parameter dependencies, a regime map was constructed, highlighting the dependence of vortex ring dynamics on $Re$ and $\delta$. Furthermore, the findings revealed that in thin liquid films, the compression of the vortex rings induced by the receding motion of the crater is the primary mechanism triggering vortex ring instabilities. This contrasts with semi-infinite domaine, where instabilities typically arise from the compression of the secondary vortex ring after orbiting towards the interior of the primary vortex ring \cite{walker1987impact,cerra1983experimental}.

Additionally, the circulation of vortex rings was measured, revealing that thinner films lead to an earlier interaction of the primary vortex ring with the wall with respect to the moment of impact. This interaction results in a rapid decrease in circulation, driven by the formation of secondary vortex rings and the development of instabilities that ultimately lead to the breakdown of the vortex rings. Moreover, an empirical model was developed to predict the temporal evolution of the total circulation, accounting for both the generation and decay phases of the vortex rings. Finally, the number of fingers resulting from the vortex ring instability was found to increase linearly with $Re$ and to depend inversely on $\delta$. This observation aligns well with the understanding that the characteristic wavelength of instabilities is directly related to the core size of the vortex ring \cite{mishra2021instability}, which varies with $\delta$.

\begin{acknowledgments}
This project is funded by the Deutsche Forschungsgemeinschaft (DFG, German Research Foundation) – project number 237267381 – TRR 150, sub-project A07 and project number 265191195 – SFB 1194, sub-project A03. \\
During the preparation of this work, the authors used ChatGPT and Grammarly to polish the English language. After using these tools, the authors reviewed and edited the content as needed and take full responsibility for the content of the published article.
\end{acknowledgments}

\section*{Data Availability Statement}

Supplementary material can be found in Citation [add Zenodo reference]. This contains all data plotted in figures, as well as the metadata of experiments. Recorded videos of droplet impacts, from which the images of figures are extracted, are also included.

\nocite{*}
\bibliography{aipsamp}

@article{cresswell1995drop,
  title={Drop-formed vortex rings—The generation of vorticity},
  author={Cresswell, RW and Morton, BR},
  journal={Physics of Fluids},
  volume={7},
  number={6},
  pages={1363--1370},
  year={1995},
  publisher={American Institute of Physics}
}

@article{lee2015origin,
  title={Origin and dynamics of vortex rings in drop splashing},
  author={Lee, Ji San and Park, Su Ji and Lee, Jun Ho and Weon, Byung Mook and Fezzaa, Kamel and Je, Jung Ho},
  journal={Nature communications},
  volume={6},
  number={1},
  pages={8187},
  year={2015},
  publisher={Nature Publishing Group UK London}
}

@article{saha2019kinematics,
  title={Kinematics of vortex ring generated by a drop upon impacting a liquid pool},
  author={Saha, Abhishek and Wei, Yanju and Tang, Xiaoyu and Law, Chung K},
  journal={Journal of Fluid Mechanics},
  volume={875},
  pages={842--853},
  year={2019},
  publisher={Cambridge University Press}
}

@article{thoraval2016vortex,
  title={Vortex-ring-induced large bubble entrainment during drop impact},
  author={Thoraval, Marie-Jean and Li, Yangfan and Thoroddsen, Sigurdur T},
  journal={Physical Review E},
  volume={93},
  number={3},
  pages={033128},
  year={2016},
  publisher={APS}
}

@article{peck1994three,
  title={The three-dimensional vortex structure of an impacting water drop},
  author={Peck, Bill and Sigurdson, Lorenz},
  journal={Physics of Fluids},
  volume={6},
  number={2},
  pages={564--576},
  year={1994},
  publisher={American Institute of Physics}
}

@article{ennayar2023lif,
  title={LIF-based quantification of the species transport during droplet impact onto thin liquid films: Species transport during droplet impact onto thin liquid films},
  author={Ennayar, Hatim and Brockmann, Philipp and Hussong, Jeanette},
  journal={Experiments in Fluids},
  volume={64},
  number={9},
  pages={148},
  year={2023},
  publisher={Springer}
}

@article{tsai1976stability,
  title={The stability of short waves on a straight vortex filament in a weak externally imposed strain field},
  author={Tsai, Chon-Yin and Widnall, Sheila E},
  journal={Journal of Fluid Mechanics},
  volume={73},
  number={4},
  pages={721--733},
  year={1976},
  publisher={Cambridge University Press}
}

@article{moore1975instability,
  title={The instability of a straight vortex filament in a strain field},
  author={Moore, Derek William and Saffman, Philip Geoffrey},
  journal={Proceedings of the Royal Society of London. A. Mathematical and Physical Sciences},
  volume={346},
  number={1646},
  pages={413--425},
  year={1975},
  publisher={The Royal Society London}
}

@article{crow1970stability,
  title={Stability theory for a pair of trailing vortices},
  author={Crow, Steven C},
  journal={AIAA journal},
  volume={8},
  number={12},
  pages={2172--2179},
  year={1970}
}

@article{archer2010instability,
  title={The instability of a vortex ring impinging on a free surface},
  author={Archer, PJ and Thomas, TG and Coleman, GN},
  journal={Journal of fluid mechanics},
  volume={642},
  pages={79--94},
  year={2010},
  publisher={Cambridge University Press}
}

@article{mishra2021instability,
  title={Instability and disintegration of vortex rings during head-on collisions and wall interactions},
  author={Mishra, Aakash and Pumir, Alain and Ostilla-M{\'o}nico, Rodolfo},
  journal={Physical Review Fluids},
  volume={6},
  number={10},
  pages={104702},
  year={2021},
  publisher={APS}
}

@article{mckeown2020turbulence,
  title={Turbulence generation through an iterative cascade of the elliptical instability},
  author={McKeown, Ryan and Ostilla-M{\'o}nico, Rodolfo and Pumir, Alain and Brenner, Michael P and Rubinstein, Shmuel M},
  journal={Science advances},
  volume={6},
  number={9},
  pages={eaaz2717},
  year={2020},
  publisher={American Association for the Advancement of Science}
}

@article{harris2012instability,
  title={Instability of secondary vortices generated by a vortex pair in ground effect},
  author={Harris, DM and Williamson, CHK},
  journal={Journal of Fluid Mechanics},
  volume={700},
  pages={148--186},
  year={2012},
  publisher={Cambridge University Press}
}

@article{cheng2010numerical,
  title={Numerical study of a vortex ring impacting a flat wall},
  author={Cheng, Ming and Lou, Jing and Luo, Li-Shi},
  journal={Journal of Fluid Mechanics},
  volume={660},
  pages={430--455},
  year={2010},
  publisher={Cambridge University Press}
}

@article{graftieaux2001combining,
  title={Combining PIV, POD and vortex identification algorithms for the study of unsteady turbulent swirling flows},
  author={Graftieaux, Laurent and Michard, Marc and Grosjean, Nathalie},
  journal={Measurement Science and technology},
  volume={12},
  number={9},
  pages={1422},
  year={2001},
  publisher={IOP Publishing}
}

@article{kissing2021delaying,
  title={Delaying leading edge vortex detachment by plasma flow control at topologically critical locations},
  author={Kissing, Johannes and Stumpf, Bastian and Kriegseis, Jochen and Hussong, Jeanette and Tropea, Cameron},
  journal={Physical Review Fluids},
  volume={6},
  number={2},
  pages={023101},
  year={2021},
  publisher={APS}
}

@article{leweke1998cooperative,
  title={Cooperative elliptic instability of a vortex pair},
  author={Leweke, Thomas and Williamson, Charles HK},
  journal={Journal of fluid mechanics},
  volume={360},
  pages={85--119},
  year={1998},
  publisher={Cambridge University Press}
}

@article{brockmann2022utilizing,
  title={Utilizing APTV to investigate the dynamics of polydisperse suspension flows beyond the dilute regime: Applying APTV to polydisperse suspensions flows},
  author={Brockmann, Philipp and Symanczyk, Christoph and Ennayar, Hatim and Hussong, Jeanette},
  journal={Experiments in Fluids},
  volume={63},
  number={8},
  pages={129},
  year={2022},
  publisher={Springer}
}

@article{seno2001,
  title={Photoionization of Rhodamine Dyes Adsorbed at the Aqueous Solution Surfaces Investigated by Synchrotron Radiation},
  author={Seno, Koichiro and Ishioka, Toshio and Harata, Akira and Hatano, Yoshihiko},
  journal={Analytical Sciences/Supplements},
  volume={17icas},
  number={ },
  pages={i1177-i1179},
  year={2001},
  doi={10.14891/analscisp.17icas.0.i1177.0}
}

@article{gendron2008diffusion,
  title={Diffusion coefficients of several rhodamine derivatives as determined by pulsed field gradient--nuclear magnetic resonance and fluorescence correlation spectroscopy},
  author={Gendron, P-O and Avaltroni, F and Wilkinson, KJ},
  journal={Journal of fluorescence},
  volume={18},
  number={6},
  pages={1093--1101},
  year={2008},
  publisher={Springer}
}

@article{ersoy2019capillary,
  title={Capillary surface wave formation and mixing of miscible liquids during droplet impact onto a liquid film},
  author={Ersoy, Nuri Erdem and Eslamian, Morteza},
  journal={Physics of Fluids},
  volume={31},
  number={1},
  pages={012107},
  year={2019},
  publisher={AIP Publishing LLC}
}

@phdthesis{geppert2019experimental,
  title={Experimental investigation of droplet wall-film interaction of binary systems},
  author={Geppert, Anne K},
  year={2019},
  school={University of Stuttgart}
}

@article{thoroddsen2003air,
  title={Air entrapment under an impacting drop},
  author={Thoroddsen, ST and Etoh, TG and Takehara, K},
  journal={Journal of Fluid Mechanics},
  volume={478},
  pages={125--134},
  year={2003},
  publisher={Cambridge University Press}
}

@article{walker1987impact,
  title={The impact of a vortex ring on a wall},
  author={Walker, JDA and Smith, Charles R and Cerra, AW and Doligalski, TL},
  journal={Journal of Fluid Mechanics},
  volume={181},
  pages={99--140},
  year={1987},
  publisher={Cambridge University Press}
}

@phdthesis{cerra1983experimental,
  title={Experimental observations of vortex ring interaction with the fluid adjacent to a surface},
  author={Cerra, Anthony William and Smith, Charles R},
  year={1983},
  school={Lehigh University}
}

@article{marmanis1996scaling,
  title={Scaling of the fingering pattern of an impacting drop},
  author={Marmanis, H and Thoroddsen, ST},
  journal={Physics of fluids},
  volume={8},
  number={6},
  pages={1344--1346},
  year={1996},
  publisher={American Institute of Physics}
}

@article{thoroddsen1998evolution,
  title={Evolution of the fingering pattern of an impacting drop},
  author={Thoroddsen, ST and Sakakibara, Jun},
  journal={Physics of fluids},
  volume={10},
  number={6},
  pages={1359--1374},
  year={1998},
  publisher={American Institute of Physics}
}

@article{cortes1995support,
  title={Support-Vector Networks},
  author={Cortes, Corinna and Vapnik, Vladimir},
  journal={Machine Learning},
  year={1995}
}

@book{awad2015efficient,
  title={Efficient learning machines: theories, concepts, and applications for engineers and system designers},
  author={Awad, Mariette and Khanna, Rahul},
  year={2015},
  publisher={Springer nature}
}

@article{stumpf2022imaging,
  title={An imaging technique for determining the volume fraction of two-component droplets of immiscible fluids},
  author={Stumpf, Bastian and Ruesch, Jonas H and Roisman, Ilia V and Tropea, Cameron and Hussong, Jeanette},
  journal={Experiments in Fluids},
  volume={63},
  number={7},
  pages={1--13},
  year={2022},
  publisher={Springer}
}

@article{verhulst1838notice,
  title={Notice sur la loi que la population suit dans son accroissement},
  author={Verhulst, Pierre-Fran{\c{c}}ois},
  journal={Correspondence mathematique et physique},
  volume={10},
  pages={113--129},
  year={1838}
}

@article{swearingen1995dynamics,
  title={Dynamics and stability of a vortex ring impacting a solid boundary},
  author={Swearingen, JD and Crouch, JD and Handler, RA},
  journal={Journal of Fluid Mechanics},
  volume={297},
  pages={1--28},
  year={1995},
  publisher={Cambridge University Press}
}

@article{roisman2008propagation,
  title={Propagation of a kinematic instability in a liquid layer: capillary and gravity effects},
  author={Roisman, Ilia V and van Hinsberg, Nils Paul and Tropea, Cam},
  journal={Physical Review E—Statistical, Nonlinear, and Soft Matter Physics},
  volume={77},
  number={4},
  pages={046305},
  year={2008},
  publisher={APS}
}

@article{saffman1979vortex,
  title={Vortex interactions},
  author={Saffman, PG and Baker, GR},
  journal={Annual Review of Fluid Mechanics},
  volume={11},
  number={1},
  pages={95--121},
  year={1979}
}

@article{yarin2006drop,
  title={Drop impact dynamics: splashing, spreading, receding, bouncing},
  author={Yarin, Alexander L and others},
  journal={Annual review of fluid mechanics},
  volume={38},
  number={1},
  pages={159--192},
  year={2006},
  publisher={Palo Alto, Calif.: Annual Reviews, 1969-}
}

@article{josserand2016drop,
  title={Drop impact on a solid surface},
  author={Josserand, Christophe and Thoroddsen, Sigurdur T},
  journal={Annual review of fluid mechanics},
  volume={48},
  number={1},
  pages={365--391},
  year={2016}
}

@article{breitenbach2018drop,
  title={From drop impact physics to spray cooling models: a critical review},
  author={Breitenbach, Jan and Roisman, Ilia V and Tropea, Cameron},
  journal={Experiments in Fluids},
  volume={59},
  number={3},
  pages={1--21},
  year={2018},
  publisher={Springer}
}

@article{schmidt2021near,
  title={Near-wall flame and flow measurements in an optically accessible SI engine},
  author={Schmidt, Marius and Ding, Carl-Philipp and Peterson, Brian and Dreizler, Andreas and B{\"o}hm, Benjamin},
  journal={Flow, Turbulence and Combustion},
  volume={106},
  number={2},
  pages={597--611},
  year={2021},
  publisher={Springer}
}

@article{pati2022numerical,
  title={Numerical and experimental investigations of the early injection process of Spray G in a constant volume chamber and an optically accessible DISI engine},
  author={Pati, Andrea and Paredi, Davide and Welch, Cooper and Schmidt, Marius and Geschwindner, Christopher and B{\"o}hm, Benjamin and Lucchini, Tommaso and D’Errico, Gianluca and Hasse, Christian},
  journal={International Journal of Engine Research},
  volume={23},
  number={12},
  pages={2073--2093},
  year={2022},
  publisher={SAGE Publications Sage UK: London, England}
}

@article{gajevic2023spray,
  title={Spray impact onto a hot solid substrate: Film boiling suppression by lubricant addition},
  author={Gajevic Joksimovic, Marija and Hussong, Jeanette and Tropea, Cameron and Roisman, Ilia V},
  journal={Frontiers in Physics},
  volume={11},
  pages={1172584},
  year={2023},
  publisher={Frontiers Media SA}
}

@article{kim2007spray,
  title={Spray cooling heat transfer: The state of the art},
  author={Kim, Jungho},
  journal={International Journal of Heat and Fluid Flow},
  volume={28},
  number={4},
  pages={753--767},
  year={2007},
  publisher={Elsevier}
}

@article{tropea1999impact,
  title={The impact of drops on walls and films},
  author={Tropea, Cameron and Marengo, Marco},
  journal={Multiphase Science and Technology},
  volume={11},
  number={1},
  year={1999},
  publisher={Begel House Inc.}
}

@article{liang2016review,
  title={Review of mass and momentum interactions during drop impact on a liquid film},
  author={Liang, Gangtao and Mudawar, Issam},
  journal={International Journal of Heat and Mass Transfer},
  volume={101},
  pages={577--599},
  year={2016},
  publisher={Elsevier}
}

@article{cossali1997impact,
  title={The impact of a single drop on a wetted solid surface},
  author={Cossali, Gianpietro E and Coghe, ALDO and Marengo, Marco},
  journal={Experiments in fluids},
  volume={22},
  number={6},
  pages={463--472},
  year={1997},
  publisher={Springer}
}

@article{maliha2022optical,
  title={Optical investigation on the interaction between a fuel-spray and an oil wetted wall with the focus on secondary droplets},
  author={Maliha, Malki and Stumpf, Bastian and Beyer, Felix and K{\"u}hnert, Marius and Kubach, Heiko and Roisman, Ilia and Hussong, Jeanette and Koch, Thomas},
  journal={International Journal of Engine Research},
  pages={14680874221095235},
  year={2022},
  publisher={SAGE Publications Sage UK: London, England}
}

@article{khan2020droplet,
  title={Droplet impact on a wavy liquid film under multi-axis lateral vibrations},
  author={Khan, Talha and Ersoy, Nuri Erdem and Eslamian, Morteza},
  journal={Experiments in Fluids},
  volume={61},
  number={8},
  pages={1--21},
  year={2020},
  publisher={Springer}
}

@article{parmentier2023drop,
  title={Drop impact on thin film: Mixing, thickness variations, and ejections},
  author={Parmentier, Justine and Terrapon, Vincent and Gilet, Tristan},
  journal={Physical Review Fluids},
  volume={8},
  number={5},
  pages={053603},
  year={2023},
  publisher={APS}
}

@article{wilkens2013vortex,
  title={Vortex ring generation during drop impact into a shallow pool},
  author={Wilkens, Andreas and Auerbach, David and van Heijst, GertJan},
  journal={arXiv preprint arXiv:1305.0403},
  year={2013}
}

@article{chu1995vortex,
  title={A vortex ring impinging on a solid plane surface—vortex structure and surface force},
  author={Chu, Chin-Chou and Wang, Chi-Tzung and Chang, Chien-Cheng},
  journal={Physics of Fluids},
  volume={7},
  number={6},
  pages={1391--1401},
  year={1995},
  publisher={American Institute of Physics}
}

@article{agbaglah2015drop,
  title={Drop impact into a deep pool: vortex shedding and jet formation},
  author={Agbaglah, Gilou and Thoraval, M-J and Thoroddsen, Sigurdur T and Zhang, Li V and Fezzaa, Kamel and Deegan, Robert D},
  journal={Journal of fluid mechanics},
  volume={764},
  pages={R1},
  year={2015},
  publisher={Cambridge University Press}
}

@article{brockmann2025enhancement,
	title={Enhancement of interfacial instabilities by solid particles during fast stretching of a liquid suspension bridge},
	author={Brockmann, Philipp and Lannert, Max and Ennayar, Hatim and Cao, Yuhao and Dong, Xulan and Zhang, Zhuocheng and Brulin, Sebastian and Roisman, Ilia V and Hussong, Jeanette},
	journal={Soft Matter},
	year={2025},
	publisher={Royal Society of Chemistry}
}

@article{eslamian2014spray,
  title={Spray-on thin film PV solar cells: advances, potentials and challenges},
  author={Eslamian, Morteza},
  journal={Coatings},
  volume={4},
  number={1},
  pages={60--84},
  year={2014},
  publisher={MDPI}
}

@article{roberts1970dynamics,
  title={Dynamics of vortex rings},
  author={Roberts, PH and Donnelly, RJ},
  journal={Physics Letters A},
  volume={31},
  number={3},
  pages={137--138},
  year={1970},
  publisher={Elsevier}
}

@book{donnelly1991quantized,
  title={Quantized vortices in helium II},
  author={Donnelly, Russell J},
  volume={2},
  year={1991},
  publisher={Cambridge University Press}
}

@article{sullivan2008dynamics,
  title={Dynamics of thin vortex rings},
  author={Sullivan, Ian S and Niemela, Joseph J and Hershberger, Robert E and Bolster, Diogo and Donnelly, Russell J},
  journal={Journal of Fluid Mechanics},
  volume={609},
  pages={319--347},
  year={2008},
  publisher={Cambridge University Press}
}

\end{document}